\documentclass[conference]{IEEEtran}
\IEEEoverridecommandlockouts
\usepackage{cite}
\pdfoutput=1
\usepackage{url}
\usepackage{amsmath,amssymb,amsfonts}
\usepackage{algorithmic}
\usepackage{graphicx}
\usepackage{textcomp}
\usepackage{xcolor}
\usepackage{amsmath,graphicx}
\usepackage{subfigure}
\usepackage{multirow}
\usepackage{amsmath, amssymb}
\usepackage{booktabs}
\usepackage{algorithm}
\usepackage{algorithmic}
\def\BibTeX{{\rm B\kern-.05em{\sc i\kern-.025em b}\kern-.08em
    T\kern-.1667em\lower.7ex\hbox{E}\kern-.125emX}}
\begin{document}

\title{CLN-VC: Text-Free Voice Conversion Based on Fine-Grained Style Control and Contrastive Learning with Negative Samples Augmentation\\
}

\author{\IEEEauthorblockN{Yimin Deng$^{1,2\ddagger}$, Xulong Zhang$^{1\ddagger}$\thanks{$^\ddagger$ Both authors have equal contributions.}, Jianzong Wang$^{1\ast}$\thanks{$^\ast$Corresponding author: Jianzong Wang (jzwang@188.com).}, Ning Cheng$^{1}$, Jing Xiao$^{1}$}
\IEEEauthorblockA{\textit{$^{1}$Ping An Technology (Shenzhen) Co., Ltd.}\\\textit{$^{2}$University of Science and Technology of China}}
}

% \author{Yimin Deng}

% \author{Yourself$^{1,2\dagger}$, Xulong Zhang$^{1\dagger}$, Jianzong Wang$^{1\ast}$\thanks{$^\dagger$These authors have equal contributions.\\$^{\ast}$Corresponding author: Ning Cheng (chengning211@pingan.com.cn)}, Ning Cheng$^{1}$, Jing Xiao$^{1}$}
% \address{$^1$Ping An Technology (Shenzhen) Co., Ltd.\\
%         $^2$University
% \author{Yimin Deng$^{1,2\dagger}$, Xulong Zhang$^{1\dagger}$, Jianzong Wang$^{1\ast}$\thanks{$^{\dagger}$These authors have equal contributions.\\$^{\ast}$Corresponding author: Jianzong Wang (jzwang@188.com).}, Ning Cheng$^{1}$, Jing Xiao$^{1}$
% \address{$^{1}$Ping An Technology (Shenzhen) Co., Ltd.\\$^{2}$University of Science and Technology of China
% }}
\maketitle

\begin{abstract}
Better disentanglement of speech representation is essential to improve the quality of voice conversion. Recently contrastive learning is applied to voice conversion successfully based on speaker labels. However, the performance of model will reduce in conversion between similar speakers. Hence, we propose an augmented negative sample selection to address the issue. Specifically, we create hard negative samples based on the proposed speaker fusion module to improve learning ability of speaker encoder. Furthermore, considering the fine-grain modeling of speaker style, we employ a reference encoder to extract fine-grained style and conduct the augmented contrastive learning on global style. The experimental results show that the proposed method outperforms previous work in voice conversion tasks.
\end{abstract}


\section{Introduction}
Voice conversion~(VC) is the process of transferring speaker identity and preserving linguistic information of speech. It has a wide range of applications in real life, such as intelligent customer service, gender anonymous, video dubbing, etc. A useful way to realize voice conversion is to disentangle speech representation and manipulate the voice characteristics like timbre and prosody to change speaker identity while preserving content.

% Achieving a change in the speaker's identity can be accomplished by modifying the speaking style. 
Nowadays, not only the naturalness but also the expressiveness of converted result play an important role in speaker style modeling. The speaking style modeling has been a subject of continuous exploration and discussion~\cite{mohamed2022self}. Early work in VC~\cite{lee2021voicemixer,yuan2021improving} uses timbre as the symbol of specific speaker. 
% AutoVC~\cite{qian2019autovc} proposed the first autoencoder framework for zero-shot VC which ensured the disentanglement of content and timbre by a bottleneck structure. 
The timbre similarity to target speaker is an important metric for the evaluation of voice conversion. The elimination of source timbre becomes necessary for the success of VC. 
Autovc~\cite{qian2019autovc} utilizes an information bottleneck to eliminate timbre while preserving content information. Instance normalization~\cite{chou2019invc} is also used to limit the leakage of timbre. Furthermore, researchers realize that timbre is not enough to fully characterize speaker style to generate convincing converted speech~\cite{deng2023PMVC}. Recently, some approaches of text-to-speech~(TTS) propose multi-scale style control for expressive speech synthesis~\cite{multiscale,audiobook,multiscale-modeling}. Multi-scale style control in TTS involves the alignment between text and prosody for better sound quality~\cite{ning2022expressive}.
However, considering not all speech datasets for VC provide text-transcript from forced-alignment~\cite{olsen2017methods}, fine-grained style modeling without text-transcript deserves further research. SpeechFlow~\cite{speechflow} is proposed for modeling pitch and rhythm to represent the speech prosody. AUTOPST\cite{autopst} proposes a down-sampling method for prosody modeling without text-transcript. 

Despite these progress in expressive voice conversion, there remains not fully explored situations where speakers have similar voices. For example, voice conversion of the same gender suffers from the similar voice ranges~\cite{padmini2022age}. In this instance, the learning ability of speaker encoder is constrained by the training process only with o reconstruction loss. To better distinct different speakers in latent space, contrastive learning is employed during this process~\cite{tang2022avqvc,tang2023vq-CL}. It benefits from the selection of appropriate positive and negative sample pairs. Typically, based on labeled speakers dataset, positive sample pair consists of speaker embeddings extracted from two utterances of the same person while negative sample pair consists of speaker embeddings extracted from those of different persons~\cite{tang2022avqvc}. Since the selection of positive samples and negative samples relies on speaker labels, the boundary of similar speakers is unclear during such training process. Recent studies point out that the robustness of contrastive learning can be improved by hard negative samples which have similar attributes and is difficult to distinct with the anchor~\cite{contrast-hard}. To improve the disentangled representation learning ability, the choice of negative samples in VC remains challenging.

To address these issues, we propose a novel Voice Conversion method based on Contrastive Learning with Negative samples augmentation and fine-grained style named ``CLN-VC''. Specially, we propose a speaker fusion module to generate augmented negative samples with labeled speakers. Since the local prosody of different utterances always varies, contrastive learning is more suitable to be applied in global features learning. Hence, we employ a reference encoder~\cite{multiscale} to extract global speaker embedding and local prosody embedding respectively. 
% As mentioned above, the VC model without text transcription is more applicable for most speech datasets. 
A content encoder based on vector quantization~(VQ)~\cite{gray1984vector} is adopted to generate content representation closed to acoustic units without text-transcript. And the alignment between content and prosody can be implemented by attention mechanism~\cite{vaswani2017attention}.
% Since the linguistic information and local prosody are both time-variant, the alignment of them becomes challenging. Recent work~\cite{multiscale} shows that models with alignment of phoneme-level content and quasi-phoneme-level prosody perform better than that with frame-level prosody. 
% To reduce such degradation, we utilize vector quantization~(VQ) for generating content features without text transcription. And the alignment can be implemented by attention mechanism. 
\begin{figure*}[htbp]
% \vspace{-2em}
  \centering
  \centerline{\includegraphics[scale=0.4]{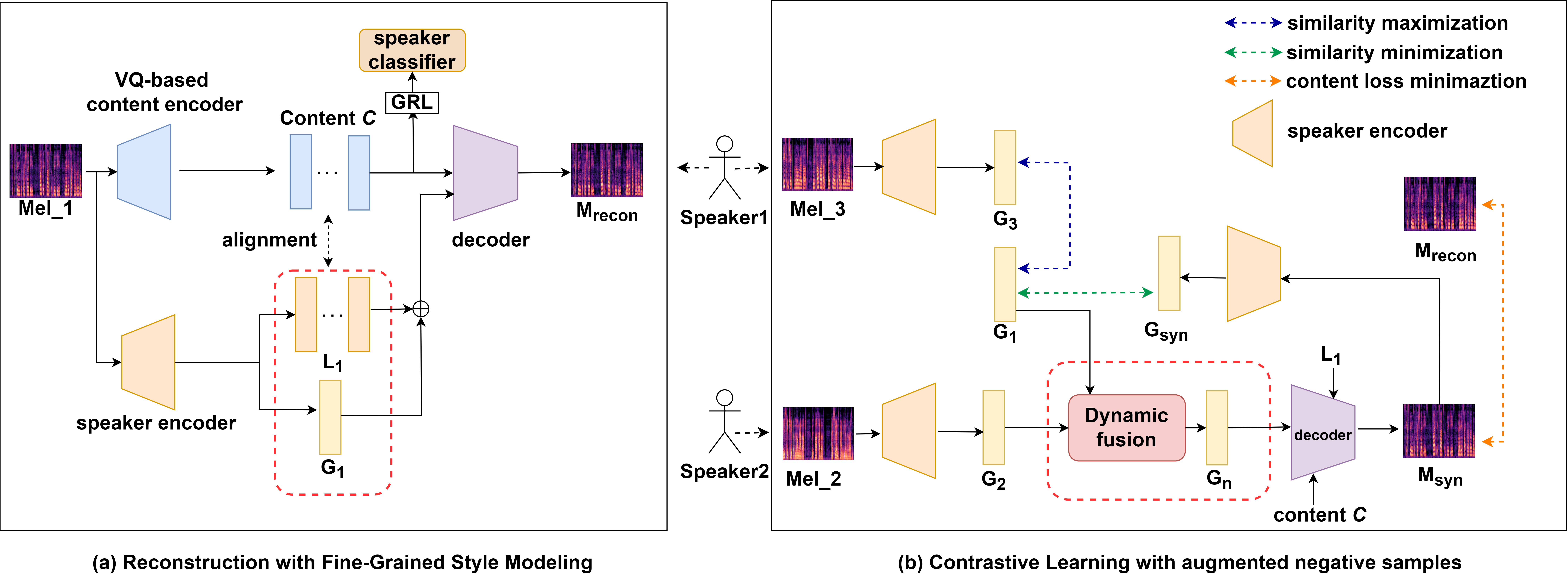}}
  \caption{ Training pipeline of proposed model. $Mel_1$ and $Mel_3$ indicate mel-spectrums from two utterances of speaker 1. $Mel_2$ means mel-spectrum from the speech of speaker 2. In sub-figure~(b), the dynamic fusion scheme is shown. $L_1$ and $G_1$ mean Local Prosody Embedding~(LPE) and Global Speaker Embedding~(GSE) from~(a). $G_2$ means GSE from the speech of speaker 2. $G_n$ is the generated GSE from dynamic fusion. $G_{sync}$ means $GSE$ from the synthesized mel-spectrum $M_{sync}$. }
  \label{fig:full_model}
  
\end{figure*}
In general, the contributions of this paper are summarized:
\begin{enumerate}
    \item We propose a speaker fusion module to generate augmented negative sample from real speakers for contrastive learning in voice conversion. With augmented negative samples in training, the performance of similar voice conversion can be improved.
    \item We integrate the fine-grained style modeling into the framework with the combination of reference encoder and VQ-based content encoder. With extracted global and local speaker style, we can apply the improved contrastive learning to the global speaker style modeling and realize expressive voice conversion with prosody modeling. 
    % \item A new method is proposed to generate augmented negative sample pairs for contrastive learning in SRL area. In training phase, another utterance of different speaker is used to extract another speaker embedding. We design two schemes of speaker fusion and consider the synthesized speech with fused speaker embedding as augmented negative sample.
    % \item Adversarial training and content consistent loss are introduced to ensure disentanglement of content-related and speaker-related information. Speaker classifier with gradient-reverse layer is used in content extraction. Minimize content consistent loss between reconstructed speech and synthesized speech to ensure quality of transfer.
\end{enumerate}

\section{Related Work}
\subsection{Voice Conversion}
A typical approach to VC tasks is to disentangle content information and speaker-related information from speech and replace the speaker representation with target. AutoVC~\cite{qian2019autovc} proposes a basic framework with autoencoders. It utilizes a bottleneck structure to encourage the learned feature to exclude speaker information so as to receive pure content information. To well represent the content information, Vector Quantization~(VQ)~\cite{VQVC} uses discrete codes from codebook which are close to acoustic units to represent content information. Text-related methods like text encoder~\cite{tang2021TGAVC}, pre-trained ASR models~\cite{zhao2022hybridASR} are also introduced to constrain the output of content encoder. However, such methods depend on annotations of datasets while the text-free VC models are more flexible.

\subsection{Contrastive Learning}
To learn speaker information, contrastive learning appears in recent VC work. The goal of contrastive learning is to encourage an encoder to encode similar data similarly and makes the encoding results of different types of data as different as possible. Its performance depends on the selection of positive sample pairs and negative ones. Early models relied on self-learning feature representations for distinguishing positive and negative samples. Supervised contrastive learning~\cite{contrast-supervised} introduces the labels from dateset as an improvement. In VC, AVQVC~\cite{tang2022avqvc} selects two utterances of the same speaker as positive pair while another utterance of different speaker as negative pair. However, some speakers in the data set have very different timbres, such as speakers of the opposite sex, but some have very similar timbres, such as speakers of the same gender. The decision boundary of the model will oscillate. Recent work has proposed multiple augmentation of original samples or added hard negative sample pairs which are hard to distinguish to improve the robustness of models. Inspired by this, we propose a novel augmentation for negative samples to improve the speaker representation ability of VC models.

\subsection{Multi-Scale Style Modeling}

Nowadays, the expressiveness of synthesized speech has aroused more and more attention. In text-to-speech area, previous work propose a reference encoder to model multi-scale speaker style including the local and global. To extract and transfer local prosody embedding, the attention-based alignment between prosody feature and content feature is important. For text-free VC models, speechsplit~\cite{speechflow} and vqmivc~\cite{vqmivc} extend the autoencoder framework adding more encoders to learn prosody. when integrating prosody modeling, the sound quality degrades much due to the lack of alignment. We propose a method which introduce reference encoder as style encoder and vq-based encoder as content encoder. Also we conduct the attention-alignment between local prosody embedding and content features with discrete codes close to acoustic units. 

\section{Methodology}
\label{sec:format}
% Given utterances of two different people, we generate augmented global speaker embedding as hard negative sample.
% The pipeline of proposed method is illustrated in Fig~\ref{fig:full_model}, which consists of two important training components: a) reconstruction with VQ-based content encoder and speaker encoder for fine-grained styles extraction; b) contrastive learning with speaker fusion module for augmented negative samples. The above components and training strategy will be described in detail as follows.

% This paper proposes a novel speech representation learning method with fine-grained style control applying contrastive learning. Voice conversion~(VC) is picked up as the downstream task of speech representation learning work. As shown in Fig~\ref{fig:full_model}, proposed method mainly consist of two  important training components: a) reconstruction with VQ-based content encoder and speaker encoder for fine-grained styles extraction; b) contrastive learning with speaker fusion module for augmented negative samples. The above components and training strategy will be described in detail as follows.

% The complete pipeline is showed as Fig \ref{fig:full_model}. The details will be described as follows.

\subsection{Disentanglement of Speech Representation}
The pipeline of CLN-VC is illustrated in Fig~\ref{fig:full_model}. The content encoder is based on vector quantization~(VQ), which discovers phone-like representation for mapping adjacent frames within the same phone into the same unit ideally. Given an input speech, a trainable codebook $CB$ is used to transfer continuous data into discrete codes. A commitment cost~\cite{VQVC} encourages each vector $Z$ of continuous feature to commit to the discrete codes and the loss is named as $L_q$.

% ~\cite{vq-wav2vec,vqvae,vq-cpc}.

% Recent research adopting vector quantization in such disentangled-based work~\cite{vq-wav2vec,vqvae,vq-cpc} has already shown that quantized result can discover phone-like representation for mapping adjacent frames within the same phone into the same unit ideally. In VQ-based content encoder~\cite{VQVC,wu2020vqvc+}, a trainable codebook $CB$ is used to transfer continuous data into discrete codes. A commitment cost is used to encourage each vector $Z$ of continuous feature to commit to the discrete codes:
% \begin{equation}
%      \mathcal{L}_{q} = \mathbb{E}[ \left\| Z-CB(Z) \right\| _{2}^{2} ]
% \end{equation}

Besides, adversarial training is used to process the output of content encoder. It's expected that the content encoder will learn as less speaker-related information as possible. As shown in Fig~\ref{fig:full_model}-a, a Gradient Reversal Layer~(GRL) is imposed before feeding the output into a speaker classifier. Therefore, the gradient is reversed by GRL before backward propagated to the content encoder. 
% Inspired by previous work~\cite{wang2021adversarially,zhao2022hybridASR}, an adversarial part with speaker classifier is designed to reduce the overlap of content and speaker information. In this process, the speaker labels are available as prediction target. It's expected that speaker classifier could not predict the labels correctly from the output of content encoder.
% As shown in Fig~\ref{fig:full_model}-a, a gradient-reverse layer is imposed before feeding the output into a speaker classifier. Therefore, the gradient of this adversarial part is reversed by GRL before backward propagated to the content encoder. 
The adversarial loss is marked as $\mathcal{L}_{adv}$ and be formulated as:
\begin{align}
     F_{\text{spk}} =& E_{\text{spk}}(\text{m}) \\
     \hat{F}_{\text{spk}} =& P_{\text{spk}}(GRL(E_{\text{con}}(\text{m}))) \\
     \mathcal{L}_{adv} =& \left\|\hat{F}_{\text{spk}}-F_{\text{spk}}\right\|_1 
\end{align}
where $E_{\text{con}}(\cdot)$ and $E_{\text{spk}}(\cdot)$ represent the output of content encoder and the ``global'' output of speaker encoder respectively. $m$ can be any mel-spectrum. $P_{\text{spk}}(\cdot)$ means the prediction made by the speaker classifier. The optimization of $\mathcal{L}_{adv}$ forces content embedding to contain speaker-related information as little as possible due to the reversal gradient imposed by GRL layer.

% \subsection{Fine-grained Style Extraction}
% As mentioned before, speaker style modeling is essential for speaker identity. For the new contrastive learning in VC task discussed following, fine-grained speaker style extraction is introduced in this section.
To learn style representation, we employ a reference encoder~\cite{multiscale} as the backbone of speaker encoder so that we can extract global speaker embedding~(GSE) and local prosody embedding~(LPE) from speech. Specially, we utilize BiGRU to learn contextual information from both forward and backward directions. All hidden-states of BiGRU form the LPE sequence. The final state of BiGRU is considered as a vector of GSE. 

The alignment of content and prosody is realized by scaled dot-product attention mechanism. First, divide LPE into two part of the same length along the feature dimension. Set content features as query and the parts of LPE as key and value respectively. Then we can get the aligned sequence~$LPE_a$:
\begin{align}
LPE_{a}&=Att.(Q,K,V)
\notag
\\&=Att.(X_C,LPE[:,:L/2],LPE[:,L/2:])
\notag
\\&=Softmax(\frac{QK^T}{\sqrt{F}})V
\end{align}
where $Att.$ indicates the attention computation, $X_C$ means content embedding, $F$ indicates the dimension of the query $X_C$. The first dimension of $LPE$ means time dimension and the second signifies feature dimension. So $L$ indicates the length of feature dimension.

Since the speaker encoder can extract fine-grained speaking style, further modification can be conducted on the global style without affecting the local style.

\subsection{Speaker Fusion for Contrastive Learning}
% With extracted global speaker embedding, contrastive learning with augmented negative samples can be well applied in voice conversion. 
% Previous work just choos
It's expected that during the training process, the model can have a good ability to distinguish speech with similar global features from different speakers. To improve this ability, the training set needs to contain samples with similar characteristics from different speakers called hard negative samples.
% It's required that the extracted global features have a certain degree of similarity in latent space. 
In constraint of current dataset with limited people, we propose two fusion schemes to create such samples.
We select GSE of one utterance of one speaker $S_1$ as the anchor sample. Take the GSE of another utterance of $S_1$ as the positive sample. The augmented negative sample will be generated by fusion with one utterance of different speaker $S_2$.
\subsubsection{Linear Fusion}
Since our goal is to reduce the distance between classes in the global feature space, it's possible to affect the global feature by adding perturbation locally in time domain. Inspired by research on speaker information modeling in UniSpeech-SAT~\cite{chen2022unispeech}, utterance mixing augmentation is introduced. With utterance mixing, the encoder will be forced to generate similar GSE. 
As shown in Fig~\ref{fig:fusion}-a, given a start position and the interval $k$, mix the utterance of $S_2$ with that of $S_1$. The mixing portion in each utterance is constrained to be less than 50\%, avoiding potential label permutation problem~\cite{stephens2000dealing}. Then extract the GSE from mixed utterance and consider it as augmented negative sample for GSE of $S_1$. 
\subsubsection{Dynamic Fusion}
\begin{figure}[htbp]
  \centering
  \includegraphics[scale=0.4]{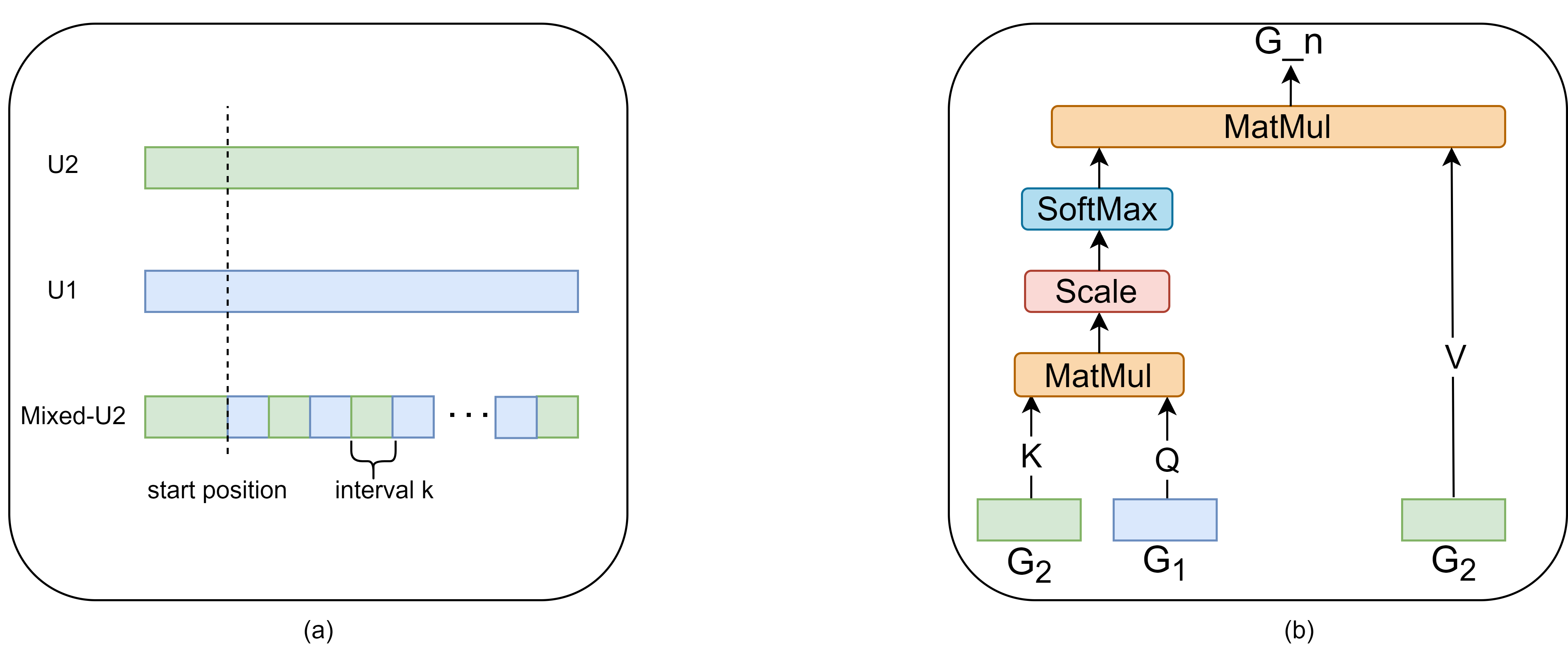}
  \caption{Speaker fusion schemes. (a) is the linear fusion. $U_1$ and $U_2$ mean utterances from speaker 1 and speaker 2 respectively. (b) is dynamic fusion.$G_1$: GSE of $S_1$, $G_2$: GSE of $S_2$, $G_n$: new GSE. Q, K ,V mean query, key and value computed with GSEs respectively.}
  \label{fig:fusion}
\end{figure}

Another fusion scheme is considered as a dynamic solution based on attention mechanism in the feature domain. Actually it's conducted on the feature space as shown in Fig~\ref{fig:full_model}-b. The hard negative sample pair should be similar and hard to distinct. We expect a channel-wise fusion method to realize the goal. To avoid generating meaningless noise, we prefer to reconstruct GSE of $S_2$ based on attention mechanism.
Usually GSE can be seen as a combination of a few areas with different attention weights. Transformation matrices $W_Q$, $W_K$, $W_V$ are used to process the vector of each GSE to conduct scaled dot-product attention. It's expected to raise the proportion of related parts and decrease that of irrelevant parts. As illustrated as Fig~\ref{fig:fusion}-b, assign different weights to attention areas according to the correlation and establish a new GSE. New speaker embedding will be used to generate hard negative sample in following step. 
\subsection{Training Strategy}
% So far, we have extracted content representation, global speaker embedding and local prosody embedding of anchor $speaker_1$ and augmented global speaker embedding in training phase. In this section, we discuss the training strategy and total loss formulation.
As shown in Fig.~\ref{fig:full_model}-(b), dynamic fusion scheme is selected in the proposed model. The necessary notations are given in Fig.~\ref{fig:full_model}.
The reconstruction task is performed on utterance $u_1$ of Speaker 1~$S_1$ with corresponding content features $C$. Reconstruction loss between $M_{recon}$ and ground truth is based on Mean Square Error~(MSE) and marked as $\mathcal{L}_{recon}$.

As said above, the improved contrastive learning is conducted on global features. Then we use the augmented GSE $G_n$ from fusion module, LPE $L_1$ and content feature $C$ from $u_1$ to synthesize new mel-spectrum $M_{sync}$. Instead of directly computing contrastive loss between $G_1$ and $G_n$, we decide to pass $M_{syn}$ through speaker encoder again and get the global feature $G_{syn}$ as hard negative sample. Because $G_{syn}$ is directly generated from speaker encoder and such consistent way seems more efficient for training. We need to increase the similarity between positive samples while decrease similarity between augmented negative samples. Cosine similarity is used as measurement: 
\begin{equation}
\begin{split}
D(G(M_{recon}), G(M_{n}))= \frac{G^T(M_{recon})G(M_{n})}{\Vert G^T(M_{recon})\Vert _2 \Vert G(M_{n})\Vert_2}
\end{split}
\end{equation}
where $D(\cdot, \cdot)$ means the cosine similarity score. $G(\cdot)$ can be any GSE extracted from input speech. $M_{n}$ represents any mel-spectrum of other speech to compose positive or negative sample pairs. The total contrastive loss for speaker representation learning can be computed as:
\begin{equation}
\mathcal{L}_{\text{sim}} = \sum^{N} (-1)^{h} D(G(M_{recon}), G(M_{n}))
\end{equation}
where h equals 1 for positive sample pairs while 0 for negative sample pairs. N indicates the number of speakers.

Besides from GSE loss, a consistent content loss $\mathcal{L}_{cc}$ between the reconstructed speech and the synthesized speech is also employed to exclude content from speaker-related information extracted by speaker encoder:
\begin{align}
    \mathcal{L}_{cc} = MSE(M_{recon}, M_{syn})
\end{align}

% To exclude content from time-variant information extracted by speaker encoder, a content consistent loss $\mathcal{L}_{cc}$ between reconstructed speech and synthesized speech is employed. 
Total loss of training process is as follows:
\begin{align}
    \label{full loss}
    \mathcal{L}(\boldsymbol{\theta_{e_c}, \theta_{e_s}, \theta_d}) = \mathcal{L}_{\text {recon}} + \alpha \mathcal{L}_{\text {sim}} + \beta \mathcal{L}_{\text {q}} + \lambda \mathcal{L}_{\text {adv}} + \gamma \mathcal{L}_{cc}
\end{align}
where $\alpha$ , $\beta$, $\lambda$ and $\gamma$ refers to the weight of $\mathcal{L}_{\text {sim}}$, $\mathcal{L}_{\text {q}}$, $\mathcal{L}_{\text {adv}}$ and $\mathcal{L}_{cc}$ respectively.  $\theta_{e_c}$, $\theta_{e_s}$ and $\theta_d$ are regularization parameters of the content encoder, speaker encoder, and decoder.
\begin{table*}[!t]
  % \setlength{\abovecaptionskip}{0pt}%    
  % \setlength{\belowcaptionskip}{10pt}%
  % \vspace{-1em}
  \caption{Comparison of different models in many-to-many VC and zero-shot VC}
  % \vspace{-1em}
  \centering
  \fontsize{8.7}{7}\selectfont
  \label{tab:Comparison}
    \begin{tabular}{ccccccc}
    \toprule
    \multirow{2}{*}{\textbf{Methods}}&
    \multicolumn{3}{c}{\textbf{Many-to-Many VC}}&\multicolumn{3}{c}{\textbf{ Zero-Shot VC}}\cr
    \cmidrule(lr){2-4} \cmidrule(lr){5-7}
    & MCD~$\downarrow$ & SMOS~$\uparrow$ & NMOS~$\uparrow$ & MCD~$\downarrow$ & SMOS~$\uparrow$ & NMOS~$\uparrow$\cr
    \midrule
    AVQVC~\cite{tang2022avqvc} & 5.31 $\pm$ 0.032 & 3.18 $\pm$ 0.041 
    & 3.31 $\pm$ 0.046 & 5.42 $\pm$ 0.018 
    & 3.12 $\pm$ 0.016 & 3.21 $\pm$ 0.035 \cr
    ClsVC~\cite{Tang2023ClsVC}& 5.24 $\pm$ 0.025 & 3.66 $\pm$ 0.022
    & 3.29 $\pm$ 0.048 & 5.36 $\pm$ 0.028 
    & 3.54 $\pm$ 0.041 & 3.26 $\pm$ 0.033 \cr
    SpeechSplit2~\cite{chan2022speechsplit2} & 5.53 $\pm$ 0.027
    & 3.35 $\pm$ 0.034 
    & 3.01 $\pm$ 0.025 & 5.89 $\pm$ 0.038
    & 3.05 $\pm$ 0.032 & 3.05 $\pm$ 0.057 \cr
    
    \midrule
    
    \textbf{CLN-VC (Linear)} & 5.11 $\pm$ 0.033 & 3.77 $\pm$ 0.018
    &\textbf{3.60 $\pm$ 0.033} & 5.33 $\pm$ 0.012 
    &3.22 $\pm$ 0.016 &3.28 $\pm$ 0.027 \cr
    
    \textbf{CLN-VC (Dynamic)} &\textbf{5.08 $\pm$ 0.012} &\textbf{3.79 $\pm$ 0.024}
    &3.58 $\pm$ 0.017 & \textbf{5.28 $\pm$ 0.015} 
    &\textbf{3.62 $\pm$ 0.026} &\textbf{3.32 $\pm$ 0.017} \cr
    \bottomrule
    \end{tabular}
    % \vspace{-1em}
\end{table*}

\section{Experiment}
In this section, we will evaluate the performance of proposed model on traditional many-to-many VC and zero-shot VC tasks. Detaily, many-to-many VC task means that in inference stage, both the selected source speaker and the target speaker are seen in training. In contrast, in zero-shot VC, both of them never appear in the training process. 
% Besides, we also empirically validate the convenience and robustness of the proposed framework. 
% We will present the audio demo and further details may be found in our implementation code.

\subsection{Datasets and Configurations}
All the objective and subjective experiments are conducted on VCTK Corpus~\cite{vctk}, a high-fidelity multi-speaker English speech corpus. It contains speech data recorded by 108 native English speakers with diverse accents for 46 hours. The entire dataset is randomly divided into 3 sets: 17262 recordings from 50 speakers for training, and other recordings from these speakers for testing. Besides, the voice of some other speakers that do not appear in training sets are used to conduct zero-shot VC experiments.

The strides of convolution blocks of speaker encoder are set as (2,1,2,1,2,2) to extract GSE and LPE. 128 was chosen as the codebook size in the content encoder. As for linear speaker fusion module, the mixing interval is set as 5. We will compare the performance of both the proposed method with linear fusion and the one with dynamic fusion with the baseline models.
% As mentioned above, we choose to integrate the dynamic fusion into the proposed method to conduct evaluations with other baselines. 
We also conduct another test to prove the efficiency of both linear fusion and dynamic fusion schemes.
The weights in Eq.(\ref{full loss}) are set to $ \alpha = 0.01, \beta = 0.1, \lambda = 0.5, \gamma = 0.5$.

AVQVC~\cite{tang2022avqvc}, ClsVC~\cite{Tang2023ClsVC}, SpeechSplit2~\cite{chan2022speechsplit2} models are chosen as the baseline models. AVQVC combines contrastive learning and VQ but without prosody modeling. ClsVC applies adversarial training while SpeechSplit2 involves fine-grained style modeling. A pre-trained Wavenet~\cite{wavenet} vocoder is used to convert all the output mel-spectrum back to the waveform.
\subsection{Comparison of VC Tasks}
% To compare the performance of different models in VC tasks under different conditions, we conduct both objective and subjective experiment on them. VC tasks can be divided into two types: many-to-many VC task and one-shot VC task. For traditional many-to-many VC task, both the selected source speaker and the target speaker of inference phase all have ever appeared in training process. In contrast, one-shot VC task is very common in real life, in which the source speaker and target speaker are unseen when training. And only require one utterance from them to convert voice.

\subsubsection{Subjective Experiment}
As an important perceptual metric, Mean Opinion Score~(MOS) test is used to evaluate the performance of parallel converted speech from different models. Natural MOS~(NMOS) describes the naturalness of results from different models. Similarity MOS~(SMOS) is used to measure the similarity between the converted voice and the ground truth which needs to concern timbre and prosody information. Both of them are higher for better. 12 volunteers~(6 males and 6 females) are asked to rate a score from 1-5 points respectively. 

As seen in Table~\ref{tab:Comparison}, CLN-VC improves the speaker similarity to target speakers and achieve a considerable degree of naturalness under different fusion schemes in many-to-many VC. In zero-shot condition, the performance of CLN-VC with linear fusion degrades evidently in similarity of voice. We attribute this to the fact that static linear transformations on limited-scale data sets are insufficient to simulate the variety of real-life timbres. While CLN-VC with dynamic fusion still performs better due to less decay of performance than other models. 
% proposed method score the best in speaker similarity and naturalness of converted speech in many-to-many VC and zero-shot VC tasks. The results supports the proposal that fine-grained style modeling improve the speaker similarity well while contrastive learning with augmented negative sample pairs 

\subsubsection{Objective Experiment}
Mel-Cepstral Distortion~(MCD) is used as objective metrics to measure the difference between the acoustic features of the transformed speech and the ground truth. The lower means the better. As shown in Table~\ref{tab:Comparison}, CLN-VC achieves lower MCD score for less distortion than baseline models.

\begin{figure}[htb]
    \centering
    \subfigure[F-F]{
        \label{same-gender conversion 1}
        \begin{minipage}[b]{0.45\linewidth}
            % \centering
            \includegraphics[width=1\textwidth]{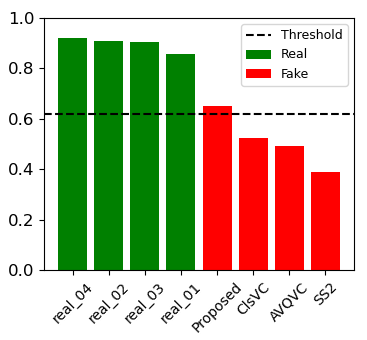}
        \end{minipage}
    } 
    \subfigure[F-M]{
         \label{cross-gender conversion 1}
        \begin{minipage}[b]{0.44\linewidth}
            % \centering
            \includegraphics[width=1\textwidth]{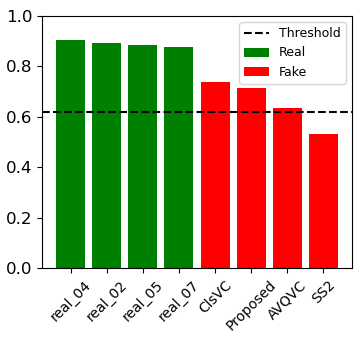}
        \end{minipage}
    }
    % \hspace{-7mm}
    % \vspace{-0.8em}
    \subfigure[M-M]{
        \label{same-gender conversion 2}
        \begin{minipage}[b]{0.45\linewidth}
            % \centering
            \includegraphics[width=1\textwidth]{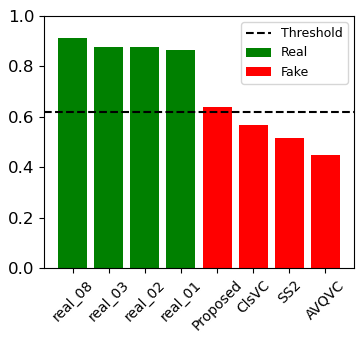}
        \end{minipage}
    }
    % \hspace{-3.6mm}
    \subfigure[M-F]{
        \label{cross-gender conversion 2}
        \begin{minipage}[b]{0.45\linewidth}
            % \centering
            \includegraphics[width=1\textwidth]{figs/M-M.png}
        \end{minipage}
    }

    \caption{Detection scores for voice conversion. F: Female; M: Male. The x-axis represents different models (Proposed: our model with dynamic fusion. SS2: SpeechSplit2) and y-axis represents the prediction score.}
    \label{fake}
    
\end{figure}

Besides, a fake speech detection test using an open-source speech detection toolkit, \textit{Resemblyzer} (\url{https://github.com/resemble-ai/Resemblyzer}) is conducted as additional evaluation in zero-shot VC condition. We prepare 10 real voices, and this toolkit automatically selects 6 of them as ”ground truth reference audios”. The rest 4 real voices and the synthetic voices from different models will be used for testing and scoring for timbre similarity. We repeat this experiment 20 times. Specially, we select the CLN-VC with dynamic fusion to take this test. As illustrated in Fig~\ref{fake}, the green groups represent the scores of real voices and the red groups represent the scores of the synthesized voice. The dash-line is the prediction threshold. Scores above the dashed line are predicted as real. With speaker fusion module, the proposed model outperforms in the same-gender VC by reaching highest scores above the dash line among fake ones. 

% We first investigate the performance of proposed model in many-to-many VC task and one-shot VC task. Different fusion schemes are proposed to promote learning ability of speaker representation with contrastive learning. In many-to-many VC task.
% We conduct a quantitative comparison between our proposed method and other baselines which is presented in Table 1. For many-to-many voice conversion, our model performs better than baselines in spectrum conversion and human perception. For one-shot voice conversion, we select a few unseen speaker from test set and input 10 utterances as content. It's showed that our model still performs well even the speakers never appear during training process which achieves a lower MCD value and higher scores in naturalness and similarity. 

\subsection{Ablation Study}

% The alignment between content and local prosody embedding deserves exploring.
In our model, several components play an important role. The evaluation of these components will be discussed as follows. The first one is the VQ technique. VQ-based content extraction is applied to mitigate the degree of quality loss. So we retrain our model with a content encoder removing VQ named ``M1``. The second one is the negative sample augmentation by speaker fusion module. To evaluate the significance of this module, we retrain the model named ``M2`` in which negative samples consists of two GSEs of utterances from different speakers after the fusion is removed. Besides, the content consistent loss $\mathcal{L}_{cc}$ is used to ensure the fidelity of content. To evaluate the importance of $\mathcal{L}_{cc}$ between reconstructed speech and another one with synthesized style, we retrain our model without $\mathcal{L}_{cc}$. We conduct the objective and subjective tests in VC of the same gender with seen speakers.
\begin{figure}[!h]
    \centering
    \subfigure[linear-based fusion]{
        \label{32-256}
        \begin{minipage}[b]{0.40\linewidth}
            % \centering
            \includegraphics[width=1\textwidth]{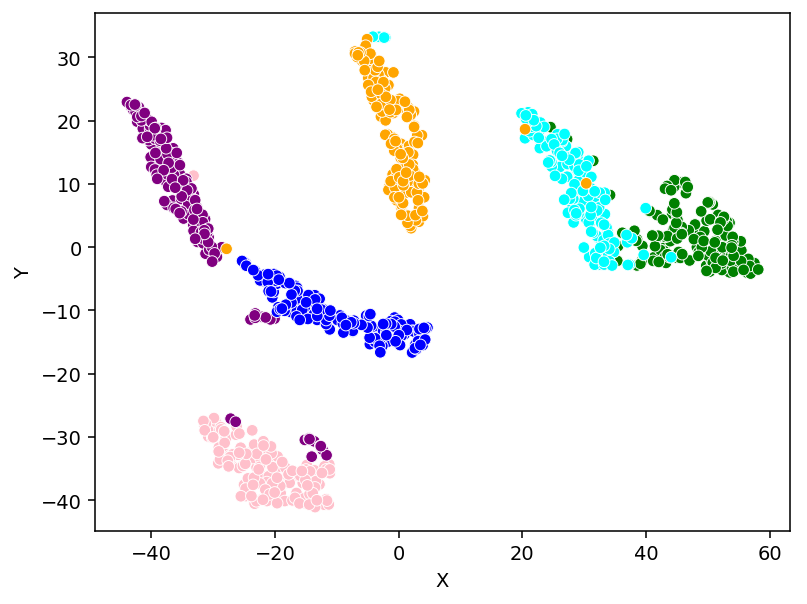}
        \end{minipage}
    }
    % \hspace{-6mm}
    \subfigure[attention-based fusion]{
         \label{32-64}
        \begin{minipage}[b]{0.50\linewidth}
            % \centering
            \includegraphics[width=1\textwidth]{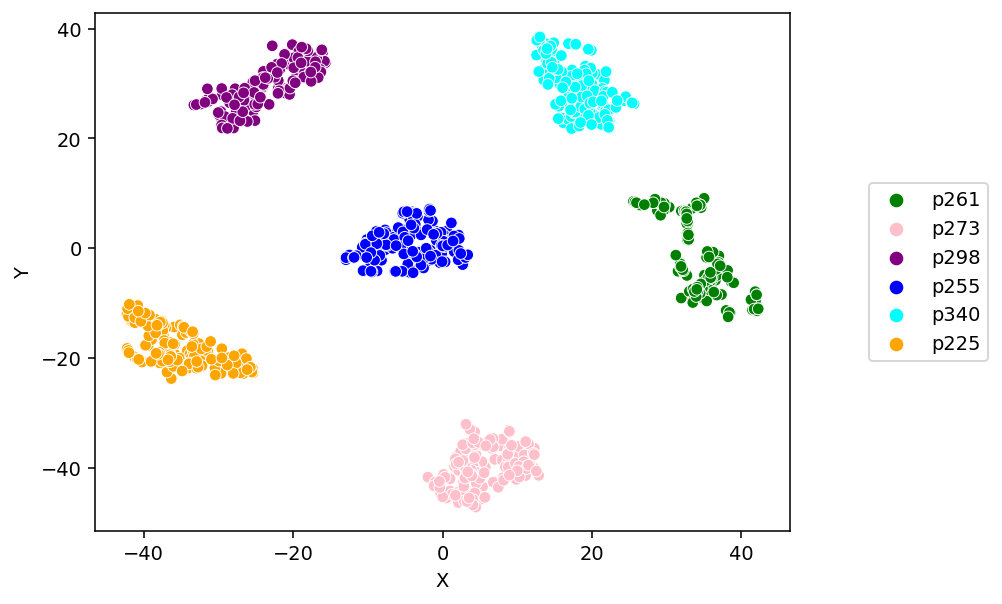}
        \end{minipage}
    }
    
    \caption{The visualization of global speaker features extracted by the models with different fusion schemes from utterances. The colors indicates different speakers.}
    \label{dim}
\end{figure}
\begin{table}[htb]
   \centering
   \caption{Results of the ablation experiments.} 
%   $\boldsymbol{C}$:Channel dimensions of the content embedding;\\
%   $\boldsymbol{S}$: Channel dimensions of the speaker embedding.}
    \label{table:2}
    
    \begin{tabular}{l c c c}
     \hline
     Method & MCD & SMOS & NMOS\\
     \hline
     CLN-VC & 3.08 $\pm$ 0.023 & 3.67 $\pm$ 0.027 & 3.45 $\pm$ 0.042\\
     w/o VQ  & 5.87 $\pm$ 0.036 & 2.58 $\pm$ 0.032 & 1.62 $\pm$ 0.029\\
     w/o fusion & 3.63 $\pm$ 0.039 & 1.52 $\pm$ 0.036 & 2.89 $\pm$ 0.035\\
     w/o $\mathcal{L}_{cc}$ & 4.71 $\pm$ 0.045 & 2.52 $\pm$ 0.026 & 2.09 $\pm$ 0.033\\
     \hline 
    \end{tabular}
\end{table}
% \vspace{-1em}

As illustrated in Table~\ref{table:2}, when removing the VQ from content encoder, the sound quality and the naturalness of converted speech degrades evidently with lower MCD and NMOS. When removing speaker fusion scheme, the performance of the retrained model degrades in the voice similarity to target with lower SMOS. As we assume above, the augmented negative samples generated from speaker fusion can improve the performance of the model in the VC task between the same gender. Besides, without $\mathcal{L}_{cc}$, the sound quality is influenced evidently, which indicates a consistent loss is a good constraint of content preservation during the training process.
\subsection{Different Fusion Schemes for Speaker Representation}
As mentioned above, we have proposed two schemes for speaker fusion to generate augmented negative samples. To further evaluate the efficiency of them for feature learning, a test is conducted to with utterances from seen speakers.
% Besides, to verify described advantages of proposed speaker fusion model, 
% A data visualization named T-distributed Stochastic Neighbor Embedding(T-SNE) is used to 
Select some utterances of them (150 utterances per speaker) as input and extract the estimated GSE $\boldsymbol{G_x}$ from speaker encoder. Then plot each hidden feature $\boldsymbol{G_x}$ in 2-D space with t-SNE as a data visualization.
% The learned speaker embeddings from different models  are mapped into 2D space using t-sne.

As shown in Fig~\ref{dim}, both of two fusion schemes can reach clear cluster patterns for speakers. However, the distance between classes is more evident in dynamic fusion scheme than that in linear fusion. Compared to linear fusion on current dataset, the VC model with dynamic fusion can distinct similar speakers and fully capture speaker-related features both in many-to-many VC and zero-shot VC. Based on these fusion schemes, more complex and effective transformation schemes on original utterances deserve further research to improve the performance of zero-shot VC model in the future.
% \vspace{-1em}

\section{Conclusion}
In this paper, we propose a novel voice conversion framework with contrastive learning and fine-grained style modeling. We use fine-grained style modeling to extract global and local speaker style and generate expressive result. Specially, we propose speaker fusion module on global speaker embedding and generate augmented negative sample pairs for contrastive learning. With augmented negative samples, we improve the performance of the model in the conversion of the same gender. Both objective and subjective experiments results demonstrate that the proposed method achieves improved performance in the naturalness of converted speech and the similarity of timbre and prosody to the target.

\section{Acknowledgement}
This paper is supported by the Key Research and Development Program of Guangdong Province under grant No.2021B0101400003. Corresponding author is Jianzong Wang from Ping An Technology (Shenzhen) Co., Ltd (jzwang@188.com).

% \clearpage
\bibliographystyle{IEEEtran.bst}
\bibliography{refs.bib}

\end{document}